\documentclass{PoS}

\title{Precision study of $K^\pm\to\pi^\pm\pi^0\pi^0$ and
$K^\pm\to\pi^\pm\pi^+\pi^-$ Dalitz plot distributions by NA48/2}

\ShortTitle{Precision study of $K^\pm\to\pi^\pm\pi^0\pi^0$ and
$K^\pm\to\pi^\pm\pi^+\pi^-$ Dalitz plot distributions by NA48/2}

\author{Evgueni Goudzovski%
        \\
        Scuola Normale Superiore and INFN, Pisa, Italy\\
        E-mail: \email{goudzovs@mail.cern.ch}}

\abstract{The NA48/2 experiment at the CERN SPS has collected an
unprecedented sample of $K^\pm\to3\pi$ decays. The high statistics
and the good resolution of the detectors allow a unique
investigation of the detailed phase space distributions of these
decays. The effects of final state pion rescattering observed in the
Dalitz plot distribution of the $K^\pm\to\pi^\pm\pi^0\pi^0$ decays
turned out to be a powerful tool for extraction of the S-wave
$\pi\pi$ scattering lengths. The large statistics also allowed a
precise measurement of the Dalitz plot slope parameters for the
$K^\pm\to3\pi^\pm$ decays.}

\FullConference{Kaon International Conference\\
        May 21-25, 2007\\
        Laboratori Nazionali di Frascati dell'INFN}

\begin{document}

\section*{Introduction}

The primary goal of the NA48/2 experiment at the CERN SPS is the
search for direct CP violation in $K^\pm\to3\pi$ decays~\cite{k3pi}.
Data have been collected in 2003--04, providing samples of $\sim
4\times 10^9$ fully reconstructed $K^\pm\to3\pi^\pm$ and $\sim10^8$
$K^\pm\to\pi^\pm\pi^0\pi^0$ decays. Surprisingly, a study of a
partial sample of $K^\pm\to\pi^\pm\pi^0\pi^0$ decays corresponding
to about 25\% of the total sample revealed an anomaly in the
$\pi^0\pi^0$ invariant mass ($M_{00}$) distribution in the region
around $M_{00} = 2m_+$, where $m_+$ is the charged pion
mass~\cite{cusp}. This anomaly, dubbed ``cusp effect'', never
observed in previous experiments, was theoretically interpreted as
an effect due mainly to the final state charge exchange scattering
process $\pi^+\pi^-\to\pi^0\pi^0$ in $K^\pm\to3\pi^\pm$ decay, and
was shown to provide a precise determination of $a_0-a_2$, the
difference between the $S$-wave $\pi\pi$ scattering lengths in the
isospin $I=0$ and $I=2$ states~\cite{ca04}. A number of theoretical
approaches to describe this process are being developed; the
original NA48/2 measurement of $a_0-a_2$ was performed in the
framework of the approach~\cite{ca05}. The current paper presents a
new step of the analysis, namely a preliminary result of a
measurement based on the full NA48/2 data sample within the same
theoretical framework.

In addition, a measurement of the Dalitz plot slopes of the
$K^\pm\to3\pi^\pm$ decay based on a partial NA48/2 data sample is
presented.

\section{NA48/2 experimental setup}

Two simultaneous $K^+$ and $K^-$ beams are produced by 400 GeV protons impinging
on a 40 cm long Be target. Particles with a
central momentum of 60 GeV/$c$ and a momentum band of $\pm3.8\%$
produced at zero angle are selected by a system of dipole magnets
forming an ``achromat'' with null total deflection, focusing
quadrupoles, muon sweepers and collimators. With $7\times10^{11}$
protons per burst of 4.5 s duration incident on the target the
positive (negative) beam flux at the entrance of the decay volume is
$3.8\times 10^7$ ($2.5\times 10^7$) particles per pulse, of which
5.7\% (4.9\%) are $K^+$ ($K^-$). The decay volume is a 114 m long
vacuum tank.

Charged particles from $K^\pm$ decays are measured by a magnetic
spectrometer consisting of four drift chambers and a large-aperture
dipole magnet located between the second and third chamber. Each
chamber has eight planes of sense wires: two horizontal, two
vertical and two along each of two orthogonal $45^\circ$ directions.
The spectrometer is located in a tank filled with helium at
atmospheric pressure and separated from the decay volume by a thin
(0.31\%$X_0$) Kevlar window. A 16 cm diameter vacuum tube centered
on the beam axis runs through the spectrometer and subsequent detectors.
Charged
particles are magnetically deflected in the horizontal plane by an
angle corresponding to a transverse momentum kick of 120~MeV/$c$.
The momentum resolution of the spectrometer is $\sigma(p)/p =
1.02\%\oplus0.044\%p$ ($p$ in GeV/$c$). The spectrometer is
followed by a scintillator hodoscope consisting of two planes
segmented into horizontal and vertical strips.

A liquid krypton calorimeter is used to reconstruct
$K^\pm\to\pi^\pm\pi^0\pi^0$ decays. It is an almost homogeneous
ionization chamber with an active volume of 7~m$^3$ of liquid
krypton, segmented transversally into 13248
projective cells of 2$\times$2~cm$^2$
by a system of Cu-Be ribbon electrodes, and with no
longitudinal segmentation. The calorimeter is 27$X_0$ thick and has
an energy resolution $\sigma(E)/E = 0.032/\sqrt{E} \oplus 0.09/E
\oplus 0.0042$ ($E$ in GeV). Spatial resolution for a single
electromagnetic shower is $\sigma_x = \sigma_y
= 0.42\%/\sqrt{E} \oplus 0.06$ cm for each transverse coordinate $x,
y$.

A detailed description of the components of the NA48 detector
can be found elsewhere~\cite{fa07}.

\section{Cusp effect and measurement of pion scattering lengths}

\begin{figure}[tb]
\begin{center}
\resizebox{0.45\textwidth}{!}{\includegraphics{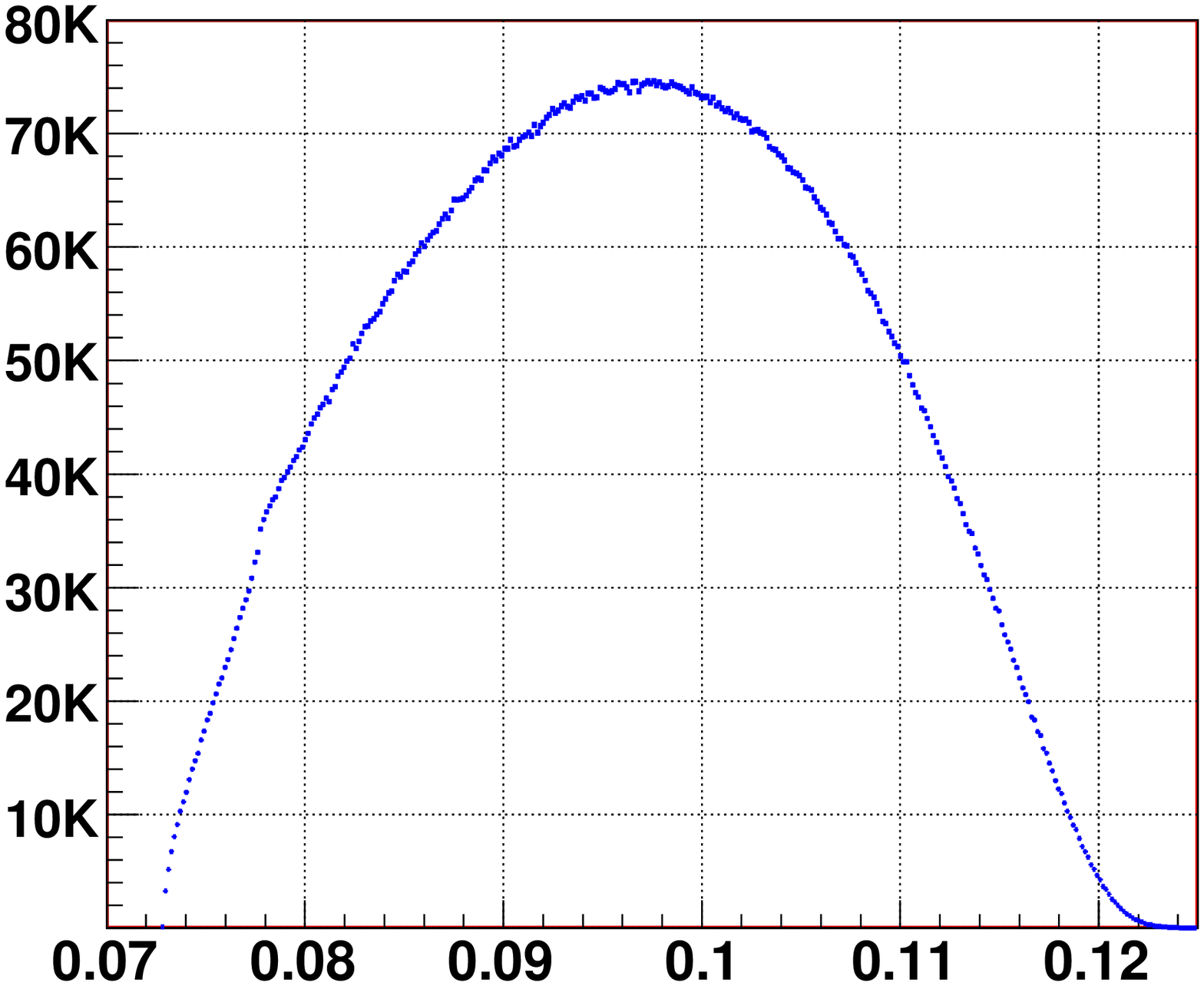}}~
\resizebox{0.455\textwidth}{!}{\includegraphics{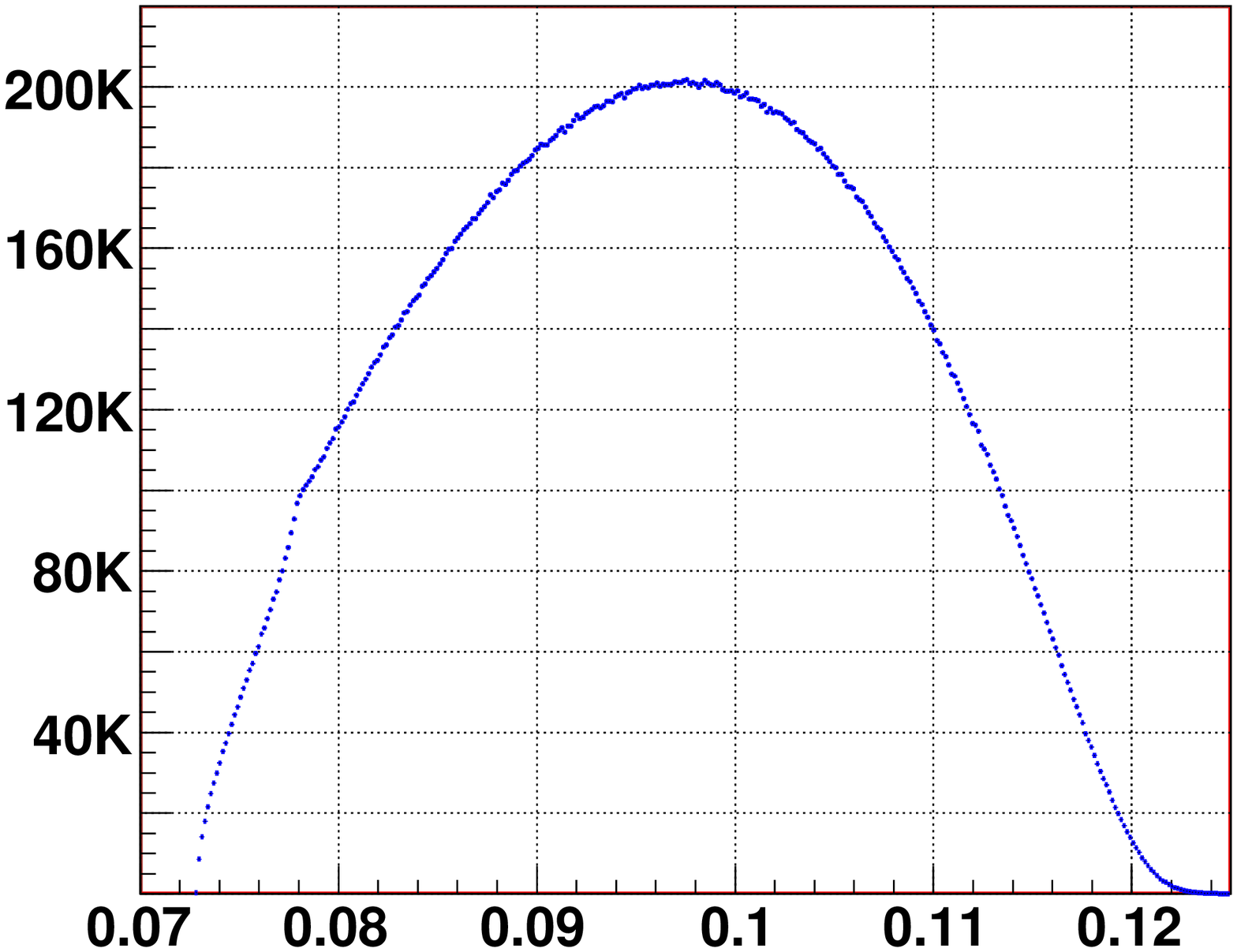}}
\put(-350,135){\Large\bf a} \put(-145,135){\Large\bf b}
\put(-70,-3){\small $M_{00}^2$, (GeV/$c^2$)$^2$}
\put(-270,-3){\small $M_{00}^2$, (GeV/$c^2$)$^2$}\\
~~~~~\resizebox{0.469\textwidth}{!}{\includegraphics{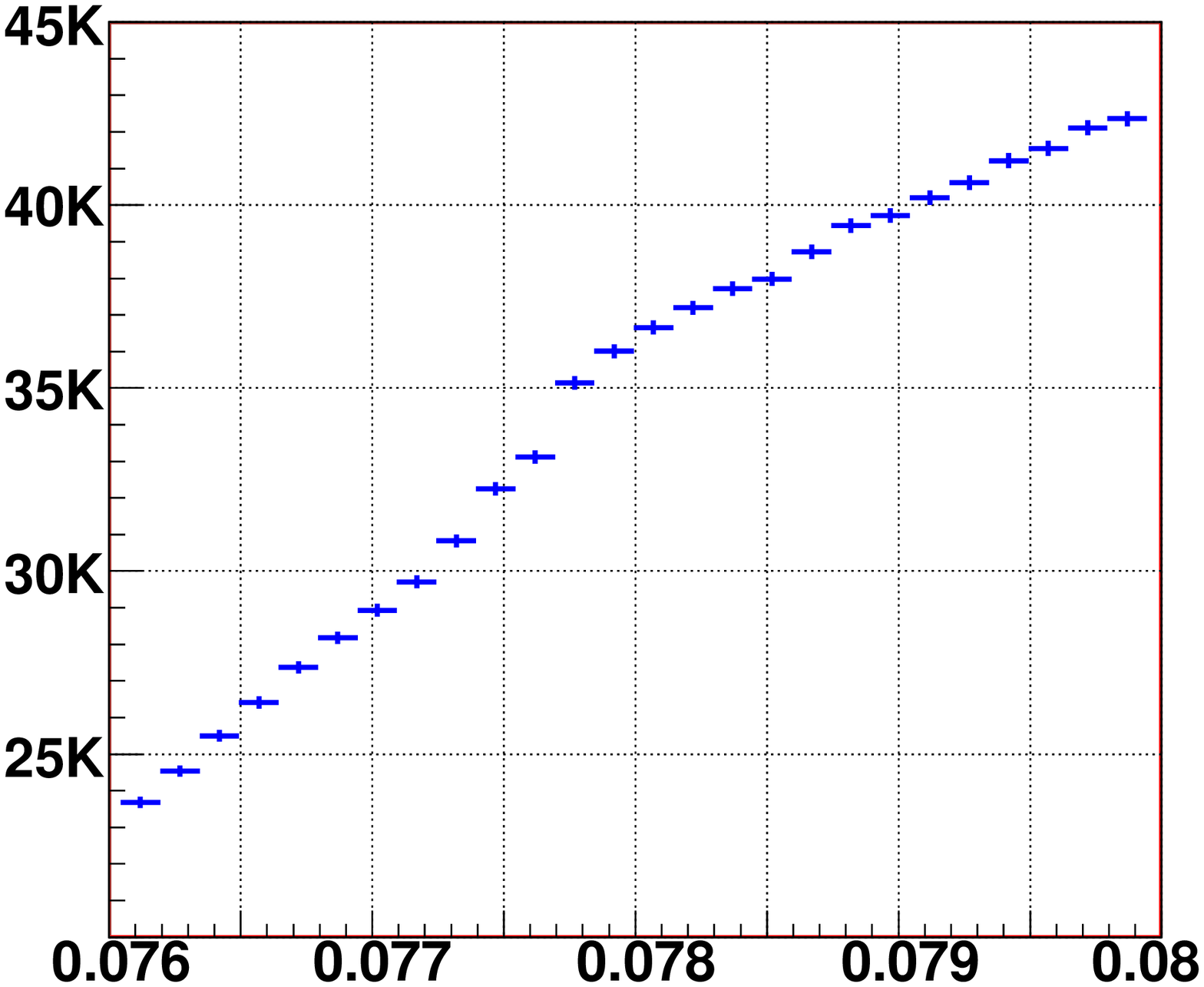}}~
\resizebox{0.46\textwidth}{0.2735\textheight}{\includegraphics{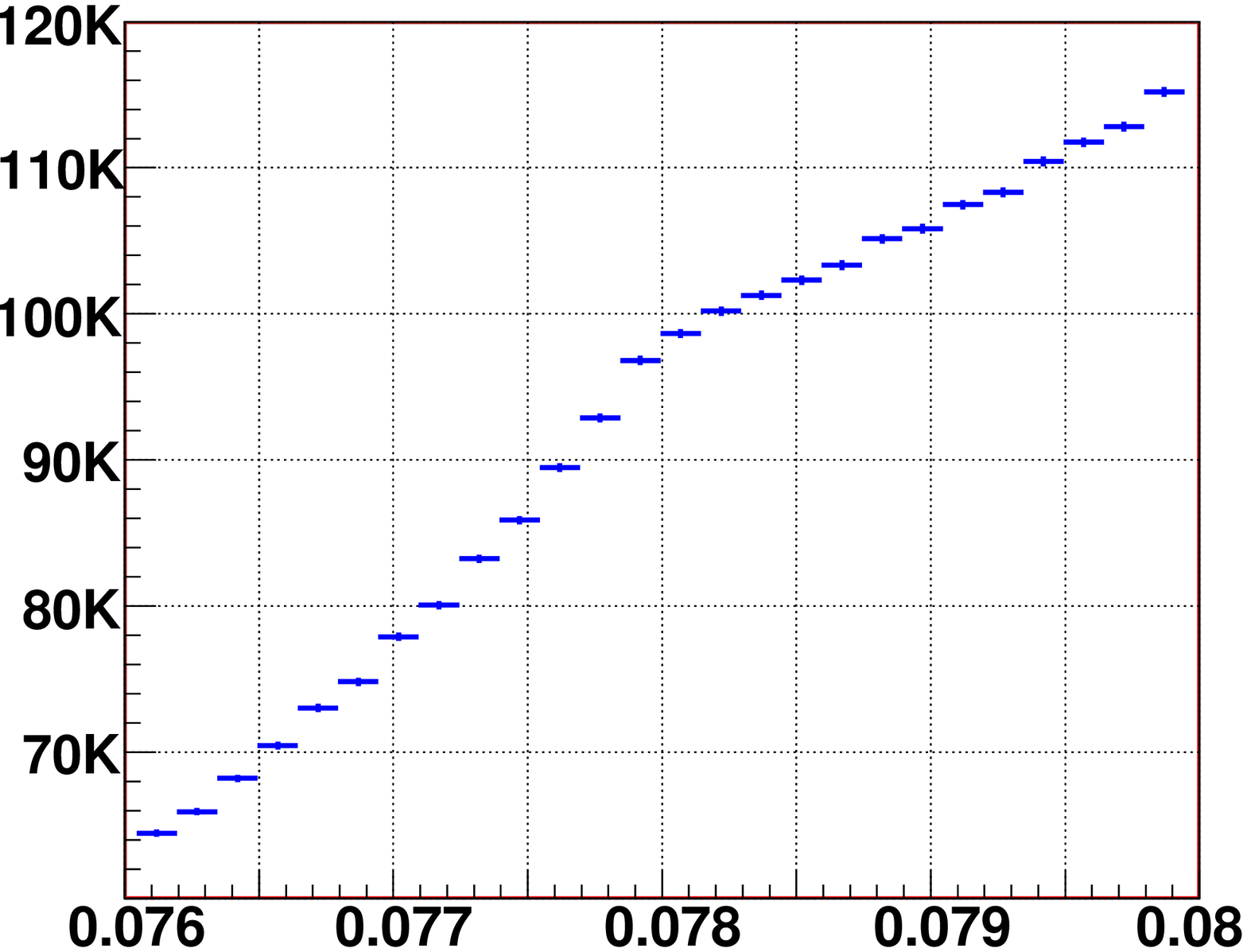}}~~~
\put(-80,-1){\small $M_{00}^2$, (GeV/$c^2$)$^2$}
\put(-280,-1){\small $M_{00}^2$, (GeV/$c^2$)$^2$}
\put(-365,135){\Large\bf c} \put(-160,135){\Large\bf d}
\end{center}
\vspace{-7mm} \caption{Reconstructed spectra of $\pi^0\pi^0$
invariant mass showing evidence for the cusp effect: the full
kinematic range for (a) 2003 data ($16.0\times10^6$ events),
(b) 2004 data ($43.6\times10^6$ events); zoomed threshold
region for (c) 2003 data, (d) 2004 data. The 2003 plots correspond
to the original discovery of the effect~\cite{cusp}, while the 2004
plots correspond to the progress with respect to the original
analysis.} \label{fig:cusp}
\end{figure}

The reconstructed spectra of $\pi^0\pi^0$ invariant mass
$M_{00}$ for 2003 and 2004 data samples (totally $59.6\times 10^9$
events) are presented in
Fig.~\ref{fig:cusp}. The change of slope at $\pi^+\pi^-$ threshold
is clearly visible. For description of this effect
the $K^\pm\to\pi^\pm\pi^0\pi^0$ amplitude is presented as a
sum of two terms:
\begin{equation}
{\cal M} = {\cal M}_0 + {\cal M}_1,
\end{equation}
where ${\cal M}_0$ is the ``unperturbed'' amplitude expressed as a
polynomial expansion in terms of the kinematic variables
$u=(s_3-s_0)/m_+^2$ and $v=(s_1-s_2)/m_+^2$, where
$s_i=(P_K-P_i)^2$, $s_0=(s_1+s_2+s_3)/3$, $P_K$ and $P_i$ are
4-momenta of kaon and pions, and $i=1,2$ correspond to the two
``even'' (i.e. identical) pions:
\begin{equation}
{\cal M}_0(u,v) = {\cal M}_0(0,0)\cdot(1 + g_0u/2 + h'u^2/2 +
k'v^2/2),
\end{equation}
and ${\cal M}_1$ is a contribution from the $K^\pm\to3\pi^\pm$ decay
amplitude ${\cal M}_+$ through $\pi^+\pi^-\to\pi^0\pi^0$ charge exchange,
which in particular simplest case of the original Cabibbo theory~\cite{ca04}
is given by
\begin{equation}
{\cal M}_1=-2a_xm_+{\cal M}_+\sqrt{1-(M_{00}/2m_+)^2}.
\end{equation}
Here, in the limit of exact isospin symmetry, $a_x=(a_0-a_2)/3$. The
amplitude ${\cal M}_1$ changes from real to imaginary at the
threshold $M_{00}=2m_+$; as a consequence it interferes
destructively with ${\cal M}_0$ below the threshold (leading to 13\%
integral depletion in this region), and adds quadratically above the
threshold.

The model used for the present measurement is based on the
formulation~\cite{ca05},
which takes into account all rescattering precesses at the one-loop and
two-loop level. In this approach
the matrix element of the $K^\pm\to\pi^\pm\pi^0\pi^0$ decay
includes a number of additional terms depending on five $S$-wave
$\pi\pi$ scattering lengths (corresponding to the processes
$\pi^+\pi^-\to\pi^0\pi^0$, $\pi^+\pi^+\to\pi^+\pi^+$,
$\pi^+\pi^-\to\pi^+\pi^-$, $\pi^+\pi^0\to\pi^+\pi^0$ and
$\pi^0\pi^0\to\pi^0\pi^0$) expressed as linear combinations of
$a_0$ and $a_2$. In addition to~\cite{ca05}, isospin
breaking effects are taken into account introducing a single
parameter $\epsilon=(m_+^2-m_0^2)/m_+^2=0.065$~\cite{ma98}.

\begin{figure}[tb]
\begin{center}
\resizebox{0.8\textwidth}{!}{\includegraphics{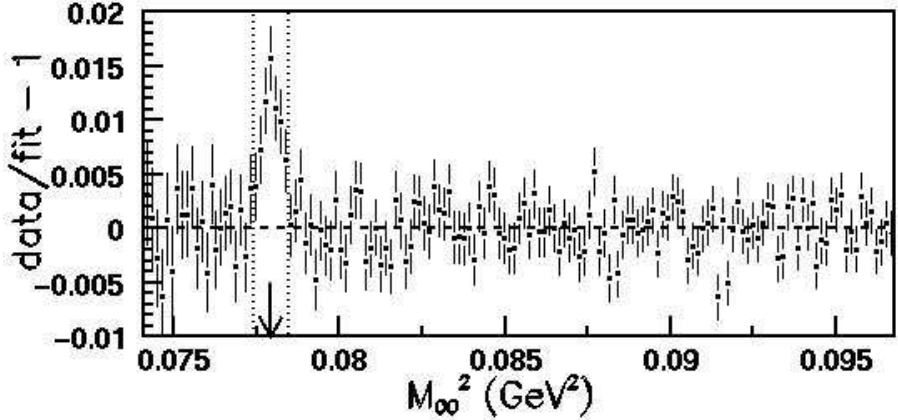}}
\end{center} \vspace{-7mm} \caption{Deviation of the data spectrum
from the fit result with statistical errors
(combined 2003+2004 data set): $\Delta=Data/Fit-1$.
Good quality of the fit and an excess of events in the region of the
threshold are demonstrated.} \label{fig:pulls}
\end{figure}

The fit to extract the scattering lengths and Dalitz plot slopes
$g_0$, $h'$ was performed in the $M_{00}$ projection of the data
using a full GEANT-based Monte Carlo simulation of the detector response.
The used rescattering model does not include radiative corrections, which
are particularly important at the threshold $M_{00}=2m_+$, and
contribute to formation of $\pi^+\pi^-$ atoms (pionium). Thus a
group of seven bins near the threshold has been excluded from the
fit. The quality of the fit ($\chi^2/{\rm NDF}=164/139$ for 2003
analysis, and $\chi^2/{\rm NDF}=119/139$ for 2004 analysis)
illustrated in Fig.~\ref{fig:pulls} shows an excess of events in
this excluded region. This excess, being interpreted as due
to pionium formation, yields the rate of pionium formation
$R=\Gamma(K^\pm\to\pi^+A_{2\pi})/\Gamma(K^\pm\to3\pi^\pm)=(1.82\pm0.21)\times 10^{-5}$,
somewhat higher than a theoretical prediction~\cite{si94}.

Measurement of the quadratic Dalitz plot slope $k'$ was performed
using the $v$ projection of the data and fixing the values of $a_0$,
$a_2$, $g_0$ and $h'$ measured by the above method. Then the fit in
$M_{00}$ projection was re-iterated to account for the measured
non-zero value of $k'$.

Systematic uncertainties due to fitting technique, trigger efficiency,
description of geometric acceptance and resolution,
calorimeter non-linearity, and simulation of showers in the calorimeter
have been evaluated.
External uncertainties due to limited experimental knowledge of
${\cal M}_+/{\cal M}_0$ at the $\pi^+\pi^-$ threshold have been
also considered. Stability checks with respect to
decay vertex position,
particle separations in the calorimeter front plane,
and kaon sign have been performed.

\section*{Results and conclusions}

The original NA48/2 measurement of the of $\pi\pi$ scattering
lengths~\cite{cusp} by exploring the cusp effect in the
$K^\pm\to\pi^\pm\pi^0\pi^0$ decay spectrum has been improved: the
full NA48/2 data sample has been used, and a more elaborate study of
systematic effects performed. The model~\cite{ca05} with isospin
breaking corrections has been used. The measured scattering lengths
are:
\begin{displaymath}
\begin{array}{rcrcrcrcr}
(a_0-a_2)m_+ &=&0.261&\pm&0.006_{\rm stat.}&\pm&0.003_{\rm syst.}&\pm&0.001_{\rm ext.}\\
a_2m_+ &=&-0.037&\pm&0.013_{\rm stat.}&\pm&0.009_{\rm
syst.}&\pm&0.002_{\rm ext.}
\end{array}
\end{displaymath}
The external uncertainties are due to the limited knowledge of
$\Gamma(K^\pm\to\pi^\pm\pi^0\pi^0)/\Gamma(K^\pm\to3\pi^\pm)$.
Moreover, an uncertainty $m_+\delta(a_0-a_2)=0.013$ has to be
attributed to the result due to precision of the theoretical model.
The Dalitz plot slopes corresponding to the used model are found to
be
\begin{displaymath}
\begin{array}{rcrcrcr}
g_0&=&0.649&\pm&0.003_{\rm stat.}&\pm&0.004_{\rm syst.}\\
h'&=&-0.047&\pm&0.007_{\rm stat.}&\pm&0.005_{\rm syst.}\\
k'&=&-0.0097&\pm&0.0003_{\rm stat.}&\pm&0.0008_{\rm syst.}
\end{array}
\end{displaymath}
In addition, a measurement of the Dalitz plot slopes of the PDG
parameterization~\cite{pdg} of the $K^\pm\to3\pi^\pm$ decay with a sample of
$4.71\times 10^8$ fully reconstructed events yielded the following results:
\begin{displaymath}
\begin{array}{rcrcrcr}
g&=&-0.21134&\pm&0.00013_{\rm stat.}&\pm&0.00010_{\rm syst.}\\
h&=&0.01848&\pm&0.00022_{\rm stat.}&\pm&0.00033_{\rm syst.}\\
k&=&-0.00463&\pm&0.00007_{\rm stat.}&\pm&0.00012_{\rm syst.}
\end{array}
\end{displaymath}
This measurement is described in detail in~\cite{slopes}. The
results are compatible with the world average, and demonstrate the
validity of the conventional parameterization at the new level of
precision.

\end{document}